%% file: Koopman.tex
\documentclass[10pt,fleqn]{article}

\usepackage{authblk}

\input{macros}

\newcommand{\Gba}{{\bar{G}}}
\newcommand{\Cb}{{\bar{C}}}

\title{A Short Introduction to the Koopman Representation of Dynamical Systems} 
\author{Bassam Bamieh}
\affil{\small Dept. of Mechanical Engineering, \\ University of California at Santa Barbara, \\{\em bamieh@ucsb.edu}} 
\date{}

\begin{document} 

\maketitle

\begin{abstract} 
The Koopman representation is an infinite dimensional linear representation of linear or nonlinear dynamical systems. It represents the dynamics of output maps (aka observables), which are functions on the state space whose evaluation is interpreted as an output. 
Conceptually simple derivations and commentary on the Koopman representation are given. We emphasize an important duality between initial conditions and output maps of the original system, and those of the Koopman representation. This duality is an important consideration when this representation is used in data-driven applications such as the Dynamic Mode Decomposition (DMD) and its variants. 
The adjoint relation between the Koopman representation and the transfer operator of mass transport is also shown.
\end{abstract}

\section{The Basic Construction}

The simplest approach  to define the Koopman representation is in a general and abstract manner using the flow map of a dynamical system. This conceptually simple approach clarifies some of the properties of the representation without getting sidetracked by the details of the underlying differential or difference equations, or modal and spectral decompositions. Those should be introduced {\em after} the basic features and properties of the representation are established. 

Consider a continuous (or discrete) time  dynamical system written abstractly  
\be
	\begin{array}{rl}
	\dot{x}_t  = &  f\left( x_t \right) , 			\\ 
	\mbox{or} ~~x_{t+1}  =&   f\left( x_t \right),	 
	\end{array}
	\hspace{4em}  x_0 = \xb,
	\hspace{4em} y_t ~=~\Gba(x_t)
  \label{dynsys.eq}	
\ee
where at each time, $x_t \in \bX$, the state space, $y_t\in\bY$, the output space, the mapping $f:\bX\rightarrow\bX$ is a vector field that generates the dynamics (or the one-step iteration in the discrete-time case), and the mapping $\Gba:\bX\rightarrow\bY$ is the output (i.e. ``readout'') mapping if the state is not directly observed, but only through the output variables $y_t$. If the state is directly observed, then the mapping $\Gba$ is simply the identity mapping. 
The spaces $\bX$ and $\bY$ 
 might be in $\R^n$ for finite vector states, or some function space when the states and outputs are spatial fields. 
 
In other treatments of the Koopman representation, the output equation is typically ignored. A point I would like to make here is that it crucial to include it even if the state is directly observed. The reason for denoting the output mapping $\Gba$ with an overbar notation will become clear shortly. 
 
 If this equation represents a well-posed dynamical system,  then there is a family (parameterized by $t$)
 of {\em  flow maps }
$\cF_t:\bX\rightarrow\bX$ such that 
\be
	x_t ~=~ \cF_{t} \left(\xb \right) , 
  \label{flow_semigroup.eq}
\ee
which map the initial condition $\xb$ to the solution $x_t$ at time $t$. The family 
 $\lcb \cF_t\rcb$ satisfies the semigroup property $\cF_{t_1+t_2} ~=~ \cF_{t_1} \circ \cF_{t_2}$, where $\circ$ is function composition. 
%
The evolution of the output starting from any initial condition $\xb$ is then given by applying the output mapping to the  state
\be
	y_t ~=~ \Gba \big( \cF_t \left( \xb \right) \big) .
  \label{KeyIdea.eq}	
\ee

\begin{figure} 
    	\centering
		\includegraphics[width=.9\textwidth]{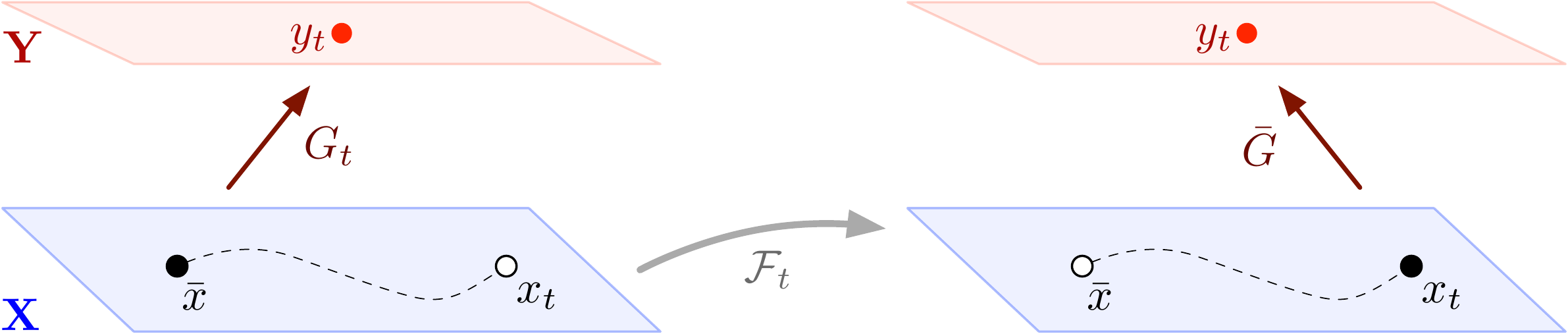} 
		
	 \mycaption{The Koopman representation is based on the idea of a {\em pullback} of one map by another. Here, the output 
	 	at time $t$ is $y_t = \Gba\big(x(t)\big) = \Gba\big( \cF_t (\xb) \big)$, 
		where $\cF_t$ is the flow map of a dynamical system, which starts from initial 
		condition $\xb$ and has output map $\Gba$. The pullback of $\Gba$ by $\cF_t$ is defined by $G_t(\xb):=\Gba\big( \cF_t(\xb)\big)$, 
		and therefore produces the same output $y_t = G_t(\xb)$ when acting as a map on the initial state. 
		When the output space $\bY$ is a vector space,  the pullback operation on $\Gba$ is alway a linear operator regardless of the 
		map $\cF_t$. } 
      \label{pullback.fig}
    \end{figure}

The key idea of the Koopman representation is to ``flip the roles'' of $\Gba$ and $\xb$ in the above equation, i.e. regard {\em evaluation at $\xb$} as an output map,  regard $\Gba$ as an initial state, and evolve $G$ rather than $x$ in time. 
Specifically, define the operator family $\{\cK_t \}$ by 
\be
	\big( \cK_t G \big) (x) ~:=~ G\big( \cF_t \left( x \right) \big) . 
  \label{KoopDef.eq}
\ee
For each $t$, the operator $\cK_t$ acts on the space $\cM(\bX,\bY)$ of all observation maps $G:\bX\rightarrow\bY$. 
Note that $\cK_t G $ is the {\em pullback} of $\Gba$ by $\cF_t$. This is illustrated in Figure~\ref{pullback.fig}. 

We now use the operator family $\lcb \cK_t \rcb$ to propagate the initial output map $\Gba$ and generate a ``trajectory'' of output maps 
\[
	G(t) ~:=~ \cK_t \big( \Gba \big) . 
\]
The original output~\req{KeyIdea} can now be obtained by simply observing  
\[
	y_t ~=~ \Gba \big( \cF_t ( \xb) \big)  ~=~ \big( \cK_t \Gba \big) (\xb) 
		~=~ \big( G_t \big) (\xb)  ~=:~ \cS_{\xb} \big( G_t \big) 
\]	
where we used~\req{KoopDef} in the second equality, and defined the {\em sampling  operator} $\cS_{\xb}\big(G\big) := G(\xb)$ which ``samples'' the
 map $G_t$ at the point $\xb$ in the state space. 
 With this construction, we have two different evolutions and output maps that {\em produce the same output $y$ }
 \[
 	\mbox{original system} \left\{ 
 	\begin{array}{rl} 
		x_t &= \cF_t(\xb) ,  \\	
		y_t &= \Gba(x_t) , 				
 	\end{array}			\right. 
	\hspace{6em} 
	\mbox{Koopman representation} \left\{
	\begin{array}{rl} 
		 G_t &= \cK_t(\Gba) 	,		\\ 
		 y(t) &= \cS_\xb \lb G_t \rb .
	\end{array}			\right. 
\]
It is in this sense that the Koopman representation is a ``representation'' of the original dynamical system. Given any trajectory 
of the original system, we can produce the same exact trajectory as an output of the Koopman system provided we start it with initial condition $\Gba$ 
and use the output map $\cS_\xb$. 

 Note  the duality of the roles of initial conditions and output maps, which are ``flipped'' between the two representations.  
  The output map $\Gba$ of the original system becomes the
  initial condition  of the Koopman evolution (which is the reason we denoted it with an overbar earlier), while the output
 map $\cS_\xb$ of the Koopman representation is parametrized by the initial conditions of the original system. 
It is important to note that even if the state is directly observed in the original dynamical system (i.e. the map $\Gba$ is the 
identity), as the Koopman state $\lcb G_t \rcb$ evolves forward in time, $G_t$ will typically not be the identity for $t>0$. 

If $\bY$ is a vector 
space, 
then $\cM(\bX,\bY)$ is endowed with a vector space structure, and it follows simply from the definition that the family $\{\cK_t\}$ is a semigroup of {\em linear} operators. Indeed 
\begin{itemize} 
	\item For each $t$, $\cK_t$ is a linear operator
		\[
		\begin{split}
			 &\lb \rule{0em}{1em} \cK_t \big( \alpha_1 G_1+ \alpha_2 G_2 \big) \rb (x) 
			~\stackrel{1}{=}~ 
			 \big( \alpha_1 G_1+ \alpha_2 G_2 \big) \big(  \cF_t ( x ) \big) 			\\
			 &\stackrel{2}{=}~
			 \alpha_1 G_1 \big( \cF_t \left( x \right) \big) 
			 + \alpha_2 G_2 \big( \cF_t \left( x \right) \big) 
			 ~\stackrel{3}{=}~ 
			 \alpha_1 \left( \cK_t G_1 \right) (x)
			 +  \alpha_2 \left( \cK_t G_2 \right) (x) ,
		\end{split}
		\]
		where $\stackrel{1}{=}$ and $\stackrel{3}{=}$ follow from the definition~\req{KoopDef}, and $\stackrel{2}{=}$
		follows from the definition of the sum of two functions that take values in a vector space. 
	\item Semigroup property: 
		\[
		\begin{split}
			\lb \cK_{t_1+t_2} G \rb (x) &=~ G\lb \cF_{t_1+t_2} (x) \rb 
			~=~ G \lb \cF_{t_2} \lb \cF_{t_1} \lb x \rb \rb \rb 			\\
			&=~ \lb \cK_{t_2} G \rb \lb \cF_{t_1} \lb x \rb \rb 
			~=~ \lb \cK_{t_1} \cK_{t_2} G \rb \lb x \rb , 
		\end{split}
		\]
		which follows from the semigroup property~\req{flow_semigroup} of the flow map. 
\end{itemize} 		
Therefore, the family $\lcb \cK_t \rcb$ is a {\em semigroup of linear operators on the space $\cM(\bX,\bY)$ of all output maps $G:\bX\rightarrow\bY$}. It is also clear that  the sampling operator $\cS_{\xb}: \cM(\bX,\bY) \rightarrow \bY$ is linear  if $\bY$ is a vector space.

    \begin{table}
	 \begingroup
	\def\arraystretch{2} \setlength\tabcolsep{2em}
           	\centering
            	\begin{tabular}{|r|c|c|}		\hline
            									&	\mbox{original system} 	& 	\mbox{Koopman representation}		 \\ \hline\hline
            	{state evolutions} $\longrightarrow$		
					&	$x_t ~ = ~   \cF_t (\xb)$ 	&   	$G_t ~ = ~   \cK_t \big(\Gba\big)$		 \\ \hline
            	{output equations} 	$\longrightarrow$		
					&	$y_t ~ = ~ \Gba(x_t) $  	&    $ y_t ~ = ~ \cS_{\xb} \big( G_t \big)~=:G_t(\xb)~$   	\\ \hline
		{state space} 
					& $\bX$ 					&	$\cM(\bX,\bY)$ 								\\ \hline
		{initial condition}	
					& $\xb$					& $\Gba$										\\ \hline
            	 \end{tabular}
               \mycaption{The relations between the original dynamical system's equations and those of its Koopman representation, 
               whose state space is the set $\cM(\bX,\bY)$ of maps between the original system's state $\bX$ and output $\bY$ spaces.
               For the 
               original system, the flow map $\cF_t$ evolves an initial condition $\xb$ to the state $x_t$ at time $t$. The output $y_t$ is obtained
               from the state by the output map (observable) $\Gba$, which may be the identity map if the state is itself the output. The Koopman 
               representation evolves the map $\Gba$ forward with the linear Koopman flow $\cK_t$. The Koopman ``state'' is then a time-varying 
               output map $G_t$. The original  output is obtained from the Koopman state  by ``sampling'' it
               $y_t=G_t(\xb)=:\cS_{\xb}(G_t)$
                at the initial condition $\xb$ of the original system.  
                Thus the roles of the initial states and output maps are reversed between the two representations. 
                Both the Koopman evolution $\cK_t$ and the Koopman output operator 
                $\cS_{\xb}$ are linear.  }
         \label{comparison.table}       
	\endgroup
     \end{table}
Table~\ref{comparison.table} shows a 
 side-by-side comparison of  the relations between the original dynamical system and its Koopman representation. 
The state space of the Koopman representation is the set  $\cM(\bX,\bY)$ of all output maps. 
The original output map $\Gba$ becomes the initial condition of an evolution governed by the linear semigroup $\lcb\cK_t\rcb$. This evolution produces a trajectory of output maps $\{G_t\}$. At each time $t$, the output signal $y_t$ is given by evaluating the current Koopman state $G_t$ at $\xb$. Thus the Koopman representation has a linear state evolution as well as a linear output map, and is therefore a linear dynamical system.

\section{Differential and Difference Equations} 

As already seen, much can be deduced about the Koopman representation from general principles 
without actually writing down the differential or difference equations of the representation. None the less, it is instructive 
and simple to derive those equations. The discrete-time case is easiest and we do that first. 

For a discrete time system 
\begin{eqnarray*}
	x_{t+1} & = & f\big( x_t \big) , \hspace{4em} x_0 = \xb ,	\\
	y_t & = & \Gba \big( x_t \big) , 
\end{eqnarray*} 
the evolution $\cF$ is simply the iterates of the map $f$, and therefore $\cF_1=f$. The generator of the corresponding Koopman representations is just $K = \cK_1$. We therefore can write 
\be
	\begin{aligned}
		G_{t+1} ~& = ~  K~ G_t , &\hspace{4em}  \lb KG \rb (x)  &:= G \big( f (x)  \big),	\\
		y_t      ~& = ~  \cS_{\xb}~  G_t,  & \lb \cS_{\xb} G \rb &:= G(\xb).
   	\end{aligned} 
   \label{DiscTimeKoop.eq}
\ee
Thus the Koopman generator is just the pullback of a function $G$ by the one-step iteration map $f$. This is a time-invariant, 
discrete-time linear system.

In continuous time, one can derive a Partial Differential Equation (PDE)  that the evolving output map $\{G_t\}$ satisfies. 
Since $\lcb \cK_t \rcb$ is a semigroup of linear operators, its generator $K$ can be calculated from the derivative at $t=0$ 
which is defined in terms of the following (strong) limit
\[
	K ~=~ \lim_{t\searrow0} \frac{1}{t} \lb \cK_t - I \rb ~=~\left.  \frac{d}{dt} \cK_t \right|_{t=0}.
\]
The action of $K$ on any function $G(.)$ is then calculated as   
\[
	\big(KG\big)(x) ~=~\left.  \frac{d}{dt} \big(\cK_t G\big) (x)  \right|_{t=0} 
		=~ \left.  \frac{d}{dt} G\big(\cF_t(x) \big)  \right|_{t=0} 
		\stackrel{1}{=}~\left.  \frac{\partial G}{\partial x} \big( \cF_t (x) \big) \frac{d \cF_t }{dt} (x) \right|_{t=0} 
		\stackrel{2}{=} \frac{\partial G}{\partial x} (x) f(x) 
\]
The  equality $\stackrel{1}{=}$ follows from the chain rule, while $\stackrel{2}{=}$ follows
from the original differential equation.  

The operator $K$ is the generator of a PDE for a time-varying output map $g(x,t):=G_t(x)$ which we now write as a function 
of $x$ and $t$ 
\be
	 \pbp{}{t}g(x,t)  ~= ~   \pbp{g}{x}\big( x ,t \big) ~f\big( x \big) 		.	
 \label{KoopPDE.eq}		
\ee
Note that $\pbp{g}{x}(x,t)$ is the 
Jacobian matrix\footnote{The Jacobian of $g$ has the entry $\frac{\partial g_i}{\partial x_j}$ in the $i$'th row 
and $j$'th column.} 
of $g$ with respect to $x$, and this equation is 
vector-valued in general. 
Equation~\req{KoopPDE} can be written in compact matrix-vector notation if $g$ and $f$ are viewed as a row, rather than 
a column vector-valued functions as follows 
\begin{align}
	\pbp{}{t}	\bbm g_1 & \cdots & g_n \ebm
	&=~ 
	\bbm f_1 & \cdots & f_n \ebm 
	\bbm 
		\sfrac{\partial g_1}{\partial x_1} & \cdots & \sfrac{\partial g_n}{\partial x_1}  \\ 
		\vdots & & \vdots \\ 
		\sfrac{\partial g_1}{\partial x_n}  & \cdots & \sfrac{\partial g_n}{\partial x_n}  \\ 		
	\ebm 											\nonumber		\\
	&=~ 
	\bbm f_1 & \cdots & f_n \ebm 
	\bbm\sfrac{\partial }{\partial x_1} \\ \vdots \\ \sfrac{\partial }{\partial x_n} \ebm 
	\bbm g_1 & \cdots & g_n \ebm 
	~=~ f \pbp{}{x} ~g  ~=:~ K ~g. 						\label{KoopGen.eq}
\end{align}
Thus if we regard $f$ and $g$ as row 
vectors\footnote{This is consistent with writing a PDE in terms of differential forms.}, 
the Koopman generator can be very compactly written as $K=f\pbp{}{x}$ as above. 
When $y$ is scalar, and thus $g$ is scalar-valued, 
 there is no difference between a row and column vector representation, and $K$ can be  
written in the more common (but clumsy) notation $f\cdot \nabla$ 
\be
	K ~=~ f \cdot \nabla ~=:~ \sum_{i=1}^n  f_i(x)  \pbp{}{x_i}. 
   \label{K_clumsy.eq}
\ee
Finally we note that since $K$ is the infinitesimal generator of the semigroup $\cK_t$, we write 
 this formally as $\cK_t ~=~ e^{t K}$.

The system~\req{KoopGen} is a linear PDE of the hyperbolic type. In fact, if $f$ is a divergence free vector field, 
then this equation is precisely the advection equation with a spatially varying velocity field of $-f(x)$. The 
trajectories of the original dynamical system are the characteristic curves of this PDE in that case. 

The Koopman representation (e.g.~\req{KoopGen}) is a linear time-invariant system, so its dynamical properties are 
completely determined by the linear operator $K$. A modal (spectral) analysis of $K$ reveals all the modes of motion of the 
system. The spectrum of $K$ is typically infinite and can have both discrete and continuous parts. Many properties 
of the original dynamical system (too numerous to mention here) can be obtained from the eigenfunctions of $K$. 
Those include limit cycles and isostables~\cite{mauroy2018global}.

\section{The Elephant in the Room: The Curse of Dimensionality}

	Just like the Hamilton-Jacobi-Bellman equation of Dynamic Programming, the Carleman linearization, 
	 the forward Kolmogorov, and the Fokker-Planck
	equations, the Koopman representation suffers from the curse of dimensionality. If the state space of the 
	original system is $\R^n$, then the state space of the Koopman representation is identified with fields over
	$n$-dimensional space. If for example, a numerical grid of size $N$ is used to discretize each state in the 
	original system, then the Koopman representation involves a discretization over a grid of size $N^n$. 
	This exponential growth in complexity makes even simulating a general Koopman representation 
	impractical\footnote{This is of course in the absence any special structure or symmetries that 
	can be exploited.}  for $n$ larger than 5 or 6. The situation is similar to simulating a PDE over $n$ 
	spatial dimensions. Simulating such a system 
	in $3$-dimensional space is feasible, but still computationally taxing for most modern computers. Similar simulations 
	over say $6$-dimensional space are only perhaps possible with the most powerful computing machines available 
	today.  

	Explicit analysis of the Koopman representation is therefore typically only done for systems of dimensions 2 or 3, 
	where the modal decomposition of the Koopman representation can offer considerable insight into the dynamical 
	system's behavior. For higher dimensional systems, one way to sidestep the curse of dimensionality is to 
	use ``data-driven'' techniques. These however suffer from another significant problem which is described next.

\section{Koopman Representation from Data: System Identification} 

The  side-by-side comparison in Table~\ref{comparison.table} clarifies an important aspect of data-driven techniques that invoke the Koopman representation. In such techniques, state or output trajectories of the original system are generated through numerical simulations or experimental observations. A corollary of the existence of the  Koopman representation implies that {\em  the same trajectories can be generated by a (infinite-dimensional) linear 
system}. 
Thus linear system identification techniques can be used to model this data. 
However, note that even if the original state is fully observed (i.e. the initial output map $\Gba$ is the identity, that is $y_t=x_t$), the Koopman  state $\lcb G_t \rcb$  is never directly  measured by the output  since the operator $\cS_{\xb}$ is not the identity. 
A set of numerically or experimentally generated trajectories correspond to a particular initial condition $\xb$. In the Koopman representation, $\xb$ appears as parametrizing the output operator $\cS_{\xb}$. To consider these  trajectories as being generated by the Koopman representation is to make a particular choice of the output operator $\cS_{\xb}$ corresponding to the initial condition $\xb$ that generated those trajectories. 

The role of the output equation in~\req{DiscTimeKoop} is not fully appreciated in the literature. For each different initial condition $\xb$ of the original dynamical system, there is a different output operator $\cS_\xb$. In this context, there is in fact not just one Koopman representation, but an infinite number of such representations parameterized by the initial condition $\xb$. All the representations share the same generator $K$, but they have different output operators. If one is trying to analyze $K$ directly from its analytical description, this distinction is irrelevant, but if one is using simulation data to identify the Koopman representation, then these distinctions become important to understand.

%
Depending on the initial condition $\xb$, the full dynamics of the Koopman representation  may or may not be observable (in the standard sense of linear systems observability) with the output operator $\cS_{\xb}$. To make this point concrete, consider the discrete-time Koopman representation~\req{DiscTimeKoop}. The {\em unobservable subspace} $\bX_\rmno$ of this system is the null space of the operator 
\[
	\bX_\rmno ~:=~ \Nus{
	\bbm \cS_\xb \\ \cS_\xb K \\ \cS_\xb K^2 \\ \vdots \ebm }, 
\]
which will typically be non-trivial, and will depend on the choice of $\xb$. Let $\bX = \bX_\rmo \oplus \bX_\rmno$ be a decomposition 
of the state space into the direct sum of the unobservable subspace, and some complement $\bX_\rmo$ of it. With respect to this 
decomposition, the system equations~\req{DiscTimeKoop} are then transformed into the 
Kalman observable decomposition~\cite{hespanha2018linear}
\begin{align*} 
	\bbm G_\rmo(t+1) \\ G_\rmno(t+1) \ebm 
	&=~ 
		\bbm K_\rmo & 0 \\ K_{\rmno\rmo} & K_\rmo \ebm 
		\bbm G_\rmo(t) \\ G_\rmno(t) \ebm  	,					\\
	y(t) &=~ \bbm 	C_\rmo & 0 \ebm \bbm G_\rmo(t) \\ G_\rmno(t) \ebm ,
\end{align*} 
where $C_\rmo$ is the restriction of the operator $\cS_\xb$ to the subspace $\bX_\rmo$. 
With this decomposition, the states $G_\rmo$ and $G_\rmno$ are termed the {\em observable and unobservable states} respectively. 
This decomposition has the property that the pair $(C_\rmo,K_\rmo)$ are observable. This means the states $G_\rmo$ can be fully reconstructed
from time series of the output $y$ (thus the term {\em observable states}). 
However, the unobservable states have no effect on the output at any time. This can be 
intuitively seen from the above equations. The states $G_\rmo$ are coupled into the states $G_\rmno$, but not vice versa, and 
only the states $G_\rmo$ are observable from the output. 
If the upper-right coupling term in the above decomposition of $K$ were not zero, 
then it might be possible to reconstruct the states $G_\rmno$ from $G_\rmo$ (which 
in turn are observable from the output). The structure of the above decomposition however precludes that possibility. 
In terms of system identification, this means that {\em only the operators
$(C_\rmo,K_\rmo)$ are identifiable from output data, while the operators $K_{\rmno\rmo}$ and $K_\rmno$ are not}. 

The above implies that there  is a part of the dynamics of the Koopman representation~\req{DiscTimeKoop} that cannot be identified from output data. However, the choice of the output $y_t = \cS_\xb G_t$ does depend on the initial condition $\xb$ of the original system. This is intuitively clear if we think about identifying the Koopman representation from simulation data.
 For some systems, a choice of $\xb$ for the simulation may not lead to exploration of the full state space and the corresponding system dynamics. For such situations, one may infer that  the Koopman system $\big(\cS_{\xb}, K \big) $ is not  fully observable. 

The preceding ideas are important to understand issues and limitations in data-driven approaches such as the Dynamic Mode Decomposition (DMD) and its variants~\cite{williams2015data,schmid2010dynamic,korda2018convergence}. When such data analysis techniques are used on trajectories of non-linear dynamical 
systems, a justification  is given  that the Koopman representation guarantees that these trajectories can also be generated 
by a linear system with possibly larger (or infinite) dimensional state space. The algorithms then proceed to do what is 
essentially 
{\em linear system identification}\footnote{This is meant to provide a larger context in which to view these methods. It is not meant 
to imply that DMD-type algorithms have already appeared in the system identification literature. On the contrary, the latter literature 
has historically been concerned with low-dimensional systems with a relatively small number of inputs and outputs. DMD on the other 
hand is tailored  for high dimensional systems and trajectories generated from Computational Fluid Dynamics (CFD) models. } 
with trajectories that were generated by a nonlinear system. 

A common issue in system identification is that the trajectories used
may not be ``rich enough''\footnote{This corresponds to the {\em persistency of excitation} condition in system identification.} 
to fully characterize the entire dynamical behavior of the original system, and this corresponds 
exactly to the Koopman representation not being fully observable. A choice of a different initial condition can be made, and the
 resulting trajectories appended for use in system identification. Another portion of the state space of the original system
 can then be explored, and this corresponds to adding another output to the Koopman representation to make it more 
 observable than with only a single simulation. 
 
 For example, suppose $N$ simulations were performed from the initial 
 conditions $\xb_1,\ldots,\xb_\ssN$. If the output time series $\lcb y_i(t) \rcb$ from each simulation are stacked together synchronously in 
 time (assuming they are all of the same length), the corresponding output equation would be 
 \[
 	y(t) ~=~ \bbm y_1(t) \\ \vdots \\ y_\ssN(t) \ebm
		~=~ \bbm \cS_{\xb_1} \\ \vdots \\ \cS_{\xb_N} \ebm G(t) 
		~=:~ \cS_{\xb_1,\ldots,\xb_N} ~G(t) .
\]
The unobservable subspace of this ``bigger'' output operator $\cS_{\xb_1,\ldots,\xb_N} $ can only be smaller than the unobservable
subspaces of the individual operators $\cS_{\xb_i}$. In fact, it will be in their intersection. This implies that more of the operator 
$K$ is identifiable than with any one simulation. For the original nonlinear system, this means that the choice of multiple initial 
conditions leads to exploration of more of the state space than with any one of them. A very intuitive conclusion.

\section{Linear Dynamics} 

It is instructive to understand the representation in the case of linear time-invariant systems. 
Consider the discrete-time case for convenience. Applying~\req{DiscTimeKoop} 
\[
	\begin{array}{rclcrclcl}
	\multicolumn{3}{c}{\mbox{\em original system}} & ~~~~~~& 
			\multicolumn{3}{c}{\mbox{\em Koopman representation}}	& & \\  \\
	 x_0& =& \xb & &  C_0  & = &  \bar{C} 		
										& & \leftarrow ~\mbox{initial condition}  \\
	x_{t+1} & = &   A ~x_t & &  C_{t+1} & = &   C_t ~A		
										& & \leftarrow ~\mbox{state equation}  \\
	y_t & = & C_o~ x_t   & &    y_t & = & C_t~ \xb  
										& & \leftarrow ~\mbox{output equation} 
	\end{array}
\]
Several  observations can be made
\begin{enumerate}
	\item 
	\textsf{Finite Dimensionality:} 
		The original output map is linear (multiplication by the matrix $\Cb$), and therefore all 
		subsequent output maps are linear. This implies that they can be finitely parametrized by the 
		entries of the matrices $C_t$, i.e. the Koopman representation is finite dimensional in this
		case! In other words, even though the Koopman representation as defined is infinite 
		dimensional, the initial state, and therefore all subsequent states are linear maps, and the 
		system never evolves outside the (finite-dimensional) subspace of linear maps from $\R^n$ to the output 
		space\footnote{The following question then immediately poses itself: For what other classes
		of dynamics is the Koopman representation finite dimensional?}. 


	\item 
	\textsf{Duality:} 
		The state iteration in the Koopman representation is given by {\em right multiplication} by 
		the matrix $A$. We can therefore conclude that the modes (eigenvectors) of the Koopman 
		representation are parametrized by the eigenvectors of $A^*$. The non-zero Koopman eigenvalues are simply 
		those of $A$ (when $A$ is real, otherwise their complex conjugates). 
	\item 
	\textsf{Grammians:} 
		The observability Grammian of the original system is 
		\[
			W ~:=~ \sum_{t=0}^\infty A^{*t}\Cb^*\Cb A^{t}.
		\]
		For stable systems, it gives a quadratic form that measures the effect of any initial condition $\xb$ 
		on the $\ell^2$ norm of the corresponding output $y$  by
		\be
			\|y \|_2^2 ~=~ \xb^* W \xb ~=~   
				\xb^*  \left( \sum_{t=0}^\infty A^{*t}\Cb^*\Cb A^{t} \right)  \xb. 
		  \label{OutGramm.eq}
		\ee
		It is interpreted as a measure  of how ``observable'' the initial condition $\xb$ 
		is from the output $y$. For example, the eigenvector of $W$ corresponding to the largest 
		eigenvalue is the most observable direction in state space. 
		
		Applying the same idea to the Koopman representation, we can naturally 
		define\footnote{The definition given here is when the original dynamics are linear. 
		In the nonlinear case, the definition would be  
		$W_K ~:=~ \sum_{t=0}^\infty K^{t} \cS_{\xb} \cS_{\xb}^* K^{*t}$, where $\cS_{\xb}$ is  
		point-evaluation operator. } 
		a ``Koopman Grammian'' 
		\[
			W_K ~:=~ \sum_{t=0}^\infty A^{t}\xb \xb^* A^{*t}. 
		\]
		The $\ell^2$ norm of an output can now be written as 
		\[
			\|y \|_2^2 ~=~ \trc{\Cb W_K \Cb^*} ~=~   
				\trc{ \Cb  \left( \sum_{t=0}^\infty A^{t}\xb \xb^* A^{*t} \right)  \Cb^* }. 
		\]
		Note that this equation is simply a reshuffle of~\req{OutGramm}, but it can be given 
		an alternate (or dual) interpretation. If we regard output maps as the object to search 
		over, then $\trc{\Cb W_K \Cb^*}$ measure how observable an output map $\Cb$ 
		makes the initial condition $\xb$.  
		
		Another setting for the Koopman Grammian  
		 is a stochastic one. If $R := \expec{\xb \xb^*}$ is a given covariance 
		of the distribution of initial states, then the eigenvectors of 
		$W_K := \sum_{t=0}^\infty A^{t}R A^{*t}$ sorted in descending order 
		of the eigenvalues, give the best choices of output maps. In other words, to choose 
		the best $q$ outputs, the optimal output map matrix $\Cb$ is the one with rows made
		from the first $q$ eigenvectors of $W_K$. 
		
	\item		
	\textsf{Non-normality:} 
		If $A$ is non-normal or $\Cb$ non-unitary, then there need not 
		be a relation between the eigenvectors of $W$ and those of $A$. This statement 
		applies equally to $W_K$ and $K$. In the nonlinear case, a similar statement 
		can be made when $K$ is non-normal, this  happens when the underlying 
		dynamical maps are not measure preserving.  
		
\end{enumerate}

\section{Relation to  the Transport Representation}
	\label{FK.sec}

The {\em transport} (aka transfer) operator 
 represents the transport of a mass distribution in state space by the dynamics of the system. It also represents the propagation in time of an initial probability density function by the unforced dynamics, i.e. it is the forward Kolmogorov (equivalently, Fokker-Planck) equation with no diffusion term. It is also referred to as the Perron-Frobenius operator in the general case, or the  Liouville operator when the dynamics are Hamiltonian. 
 The transfer operator gives another linear representation of the dynamics of a nonlinear system. We will call this the {\em transport representation}, and show that it and the Koopman representations are  are adjoints of each other.

\begin{figure}[h]
	\begin{center}	\includegraphics[height=0.1\textheight]{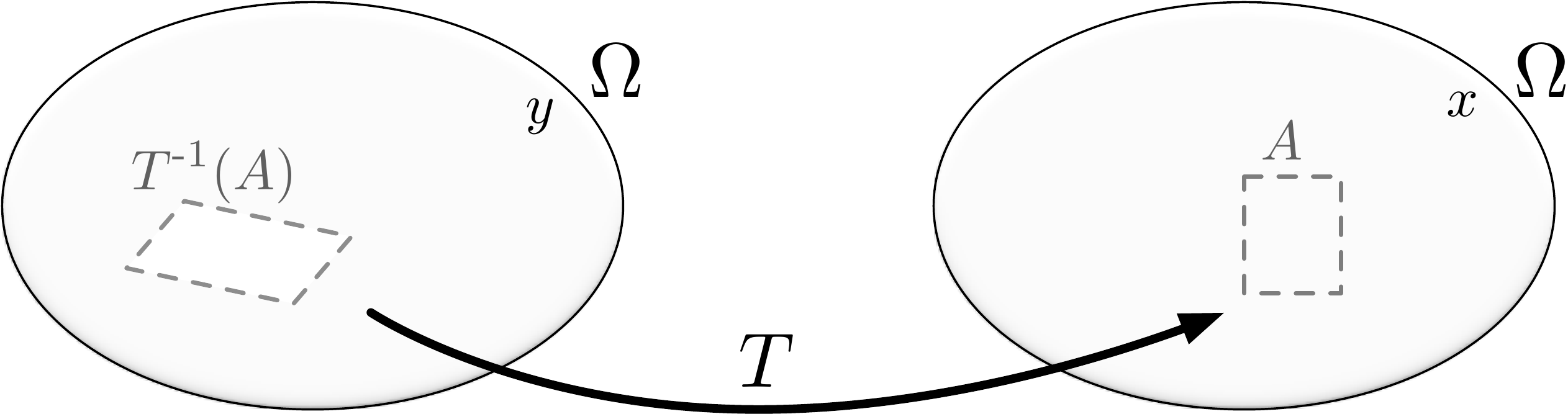}	\end{center} 
   \mycaption{Mapping of mass densities by the transformation $T$. The pre-image $T^{-1}(A)$ of the volume element $A$  contains the same mass as $A$. }
   \label{transp.fig}
\end{figure}

We first present a self contained derivation of the transport representation. 
Let $T:\Omega \longrightarrow \Omega$ 
be an invertible map defined on a subset $\Omega$ of a measure space (e.g. $\R^n$). One can think of this map 
as describing the transport of some material with non-uniform density in $\Omega$. An expression for the final density in terms of the original density and 
the transport map $T$ can be easily derived. This situation is illustrated in Figure~\ref{transp.fig}.
Let $y$ and $x$ be a coordinate system in the domain and range respectively, and let
$\phi_1$ and $\phi_2$ be the density functions before and after the transformation $T$ 
respectively. If we consider an arbitrary volume element $A$ in the range, the total 
mass of material in that volume element is given by 
\[
	\int_A \phi_2(x) ~dx ~=~ \int_{T^{-1} (A)} \phi_1(y) ~dy, 
\]
where equality follows from the fact that the material in $A$ is transported from the 
volume element $T^{-1}(A)$. Now using the transformation $x=T(y)$ and changing 
variables in the second integral yields 
\[
	\int_A \phi_2(x) ~dx ~=~ \int_{A} \phi_1(T^{-1}(x))  
					\left| \frac{\partial T^{-1}}{\partial x}(x) \right|  ~dx, 
\]
where $\left| \frac{\partial T^{-1}}{\partial x}(x) \right| $ is the determinant of the Jacobian 
of the map $T^{-1}$. Since this equality holds for any volume element, we have derived
the relation between $\phi_1$ and $\phi_2$ as
\be
	\phi_2(x) ~=~ 
	\phi_1 \left( T^{-1}(x) \right) ~ \left| \frac{\partial T^{-1}}{\partial x}(x) \right| .
   \label{perrdef1.eq}
\ee
This motivates our formal definition of the transport semi-group $\cT_t$
which acts on functions $\phi:\R^n\rightarrow \R$ that are transported by the 
flow $\cF$  of the dynamical system~\req{dynsys}
\be
	\big(\cT_t\phi\big) (x) ~:=~ 
		\phi \left( \cF_t^{-1}(x) \right) ~ \left| \frac{\partial \cF_t^{-1}}{\partial x}(x) \right| .
   \label{perrdef.eq}
\ee
We note that 
although this definition is motivated in terms of material transport, an identical 
interpretation can be carried out in terms of probability density functions.  This formula is also valid for {\em vector-valued} densities $\phi$, where each vector component of $\phi$ can be interpreted as the density of a distinct component of a material. 

%

Comparing~\req{perrdef} with~\req{KoopDef}, we see that $\cT_t$ can be thought of as a 
{\em push forward} of the function $\phi$ by the map $\cF_t$ with a ``weighting'' by the determinant 
of the Jacobian. Therefore it is not surprising that there
 is a formal mathematical connection between the transport and Koopman evolution 
semi-groups: they are adjoints with respect to the $L^2$ inner product. 
To verify this, 
start from the definitions~\req{perrdef},~\req{KoopDef}, assume for simplicity that $\Omega=\bX=\R^n$, and 
calculate for any two functions $\phi, \psi \in L^2(\R^n)$ 
\begin{eqnarray*}
	\inprod{\phi}{\cK_t\psi} & = &  
		\int_{\R^n} \phi^*(y) ~\left(\cK_t\psi\right)(y) ~dy ~=~   
		\int_{\R^n} \phi^*(y) ~\psi\left( \cF_t(y) \right) ~dy					\\
		& = &  
		\int_{\R^n} \phi^* \left( \cF^{-1}_t(x) \right) ~\psi(x)  
			~ \left| \frac{\partial \cF_t^{-1}}{\partial x}(x) \right|~dx	
		~=~ \int_{\R^n} \left( \cT_t\phi\right)^*\hspace{-0.2em}(x) ~\psi(x) ~dx 			\\
		& = &  \inprod{\cT_t\phi}{\psi},
\end{eqnarray*}
where we have used the change of variables $y=\cF_t(x)$ in the integration.
This shows that indeed $\cT_t^* = \cK_t$.


The infinitesimal generator $K=f\pbp{}{x}$  of $\cK_t$ has already been computed in~\req{KoopGen}. 
To compute the infinitesimal generator of the transport operator, it is possible 
to repeat this exercise using the definition~\req{perrdef} of $\cT_t$. 
Alternatively, we can exploit the fact that   the semi-groups $\cK_t$ and $\cT_t$ are
adjoints, which means their infinitesimal generators $K$ and $T$ are also adjoints. 
This is true under fairly mild
conditions on the semigroups, and symbolically is written as 
\[
	\cK^*_t = \left(e^{tK}\right)^* = e^{tK^*} = e^{tT}  = \cT_t . 
\]
The computation of the generator adjoint $T=K^*$ is a fairly easy 
exercise in integration by 
parts\footnote{Caution: The operator $T$ here is not the same as the transformation $T$ in~\req{perrdef1}, which was only used to 
motivate the definition of the transport semigroup~\req{perrdef}.}. 
For simplicity, we do this for the scalar case (i.e. single output, and the space of observables is thus scalar valued functions on the state space).  First expand the inner product $\inprod{\phi}{K\psi}$ by
\begin{eqnarray*}
	\inprod{\phi}{K\psi} 
	& = & 	 \int \phi^*(x) ~ \big(K\psi\big)(x) ~dx
		~=~ \sum_{i=1}^n \int \phi(x) ~f_i(x) \pbp{\psi}{x_i} ~dx		\\
	& = & 	
	-\sum_{i=1}^n \int \pbp{}{x_i} \hspace{-0.2em} \left(\rule{0em}{1em} \phi(x) f_i(x)\right)  ~\psi(x) ~dx
	~=~  \inprod{-\nabla (\phi f)}{\psi}
\end{eqnarray*}
We thus discover that 
\[
	T\phi
	~ = ~ K^*\phi
	~=~ -\nabla(\phi f).
\]	

The last expression can be written in several different forms by observing that
\begin{eqnarray*}
	\sum_{i=1}^n \pbp{}{x_i} \hspace{-0.2em} \left(\rule{0em}{1em} \phi f_i\right)
	&=&
	\sum_{i=1}^n \left(  \pbp{\phi}{x_i}  f_i   ~+~  \phi \pbp{f_i}{x_i}   \right)		\\
	&=&  \sum_{i=1}^n \left(    f_i  \pbp{}{x_i} +   \pbp{f_i}{x_i}   \right) ~\phi.
\end{eqnarray*}
The operator $T$ is therefore sometimes  written symbolically as
\[
	T ~=~ -\left( f\cdot\nabla + (\nabla f) \right). 
\]
Therefore $T$ is the sum of a differential operator $f\cdot \nabla$ and the operator of multiplication by the function $\nabla f$. 

In the case when $f$ is a divergence free vector field (i.e. $\nabla f=0$), this
expression simplifies to $T = -f\cdot\nabla$. Comparing this with~\req{K_clumsy}, we see that in the case of divergence-free $f$
\[
	K^* ~=~ T ~=~  -f\cdot \nabla ~=~    - K,  
	~~~\mbox{and}~~~  
	T^* ~=~ K ~=~ -T
\]
i.e. both operators $T$ and $K$ are skew Hermitian. This implies that the semi-groups $\cT_t=e^{tT}$ and $\cK_t=e^{tK}$ are unitary, that is, their evolutions preserve  $L^2$ norms. 

Finally, we note that the partial differential 
equation for a density function propagated by the transport operator is 
\[
	\pbp{\phi}{t}(x,t) ~=~ 
	-\sum_{i=1}^n \pbp{}{x_i} \hspace{-0.2em} \left(\rule{0em}{1em}f_i(x) \phi(x,t) \right).
\]

\bibliographystyle{plain} 
\bibliography{Koopman}

\end{document}

%% file: macros.tex
    \usepackage{amsmath,amssymb,amsfonts,amsbsy,amsthm}
    \usepackage{latexsym,graphics,graphicx,color,sectsty,mathdots,bm}
    
    \usepackage{subcaption}
    \usepackage{layout}
    \usepackage{bbm}
    \usepackage[colorlinks=true,linkcolor=blue]{hyperref}
    \usepackage{calc,cancel}
    \usepackage{enumitem}
    \usepackage{rotating}
    \usepackage[table]{xcolor}
    \usepackage{cleveref}
    \usepackage{appendix} 
    \usepackage{sidecap}
    \usepackage{amscd}

    \usepackage{arydshln}
    \usepackage{makeidx} 
    \usepackage{nicematrix}
    \usepackage{fancyhdr}
    \usepackage{nicematrix}
    \usepackage{xfrac}

\theoremstyle{plain}

\newtheorem*{summary*}{Summary}
\theoremstyle{definition}
\newtheorem{example}{Example}

\theoremstyle{remark}

\newcommand{\trc}[1]   { {\rm trace}\left( #1 \right) }

	\definecolor{bgblue}{rgb}{0.04,0.39,0.53}
	\definecolor{dblue}{rgb}{0,0.3,0.7}
	\definecolor{ddblue}{rgb}{0,0.1,0.6}
	\definecolor{ddgreen}{rgb}{0,0.25,0.05}
	\definecolor{dgreen}{rgb}{0,0.5,0.05}

\newcommand{\enma}[1]   {\ensuremath{#1}}

\newcommand{\req}[1]{(\ref{#1.eq})}

\newcommand{\beq}{\begin{equation}}
\newcommand{\eeq}{\end{equation}}
\newcommand{\beqn}{\begin{eqnarray}}
\newcommand{\eeqn}{\end{eqnarray}}
\newcommand{\beqns}{\begin{eqnarray*}}
\newcommand{\eeqns}{\end{eqnarray*}}
\newcommand{\bct}{\begin{center}}
\newcommand{\ect}{\end{center}}
\newcommand{\btmz}{\begin{itemize}}
\newcommand{\etmz}{\end{itemize}}
\newcommand{\benum}{\begin{enumerate}}
\newcommand{\eenum}{\end{enumerate}}

\newcommand{\R}{{\mathbb R}}

\newcommand{\cF}{\enma{\mathcal F}}

\newcommand{\cS}{\enma{\mathcal S}}

\newcommand{\xb}{{\bar{x}}}

\newcommand{\bbm}{\begin{bmatrix}} 
\newcommand{\ebm}{\end{bmatrix}} 

\newcommand{\bsm}{\left[ \begin{smallmatrix}} 
\newcommand{\esm}{\end{smallmatrix} \right]} 

\newcommand{\bbNm}{\begin{bNiceMatrix}} 				
\newcommand{\ebNm}{\end{bNiceMatrix}} 
\newcommand{\bNA}[1]{ \left[ \begin{NiceArray}{#1} } 		
\newcommand{\eNA}{ \end{NiceArray} \right] }

\newcommand{\lb}{\left(}
\newcommand{\rb}{\right)}
\newcommand{\lcb}{\left\{}
\newcommand{\rcb}{\right\}}

\newcommand{\cM}{{\mathcal M}}

\newcommand{\bX}{\mathbf{X}}
\newcommand{\bY}{\mathbf{Y}}

\newcommand{\cK}{{\mathcal K}}
\newcommand{\cT}{{\mathcal T}}

\newcommand{\be}{\begin{equation}}
\newcommand{\ee}{\end{equation}}

\newcommand{\cplxs}{ C\kern -.35em \rule{0.03 em}{.7 ex}~   }

\def\complex{\hbox{C\kern -.45em \rule{0.03 em}{1.5 ex}}~}

\newcommand{\bi}{\begin{itemize}}
\newcommand{\ei}{\end{itemize}}
\newcommand{\ben}{\begin{enumerate}}
\newcommand{\een}{\end{enumerate}}

\newcommand{\pbp}[2]{\frac{\partial #1}{\partial #2}}

\newcommand{\expec}[1]{{\mathcal E}\left\{ #1 \right\}}

\newcommand{\inprod}[2]{\left< #1 \boldsymbol{,} #2 \right>}

\newcommand{\bseq}{\begin{subequations}}
\newcommand{\eseq}{\end{subequations}}

\newcommand{\ba}{\begin{array}}
\newcommand{\ea}{\end{array}}

\newcommand{\mycaption}[1]{\caption{\footnotesize #1}}

\definecolor{dred}{rgb}{.8,0,0}

\def\clap#1{\hbox to 0pt{\hss#1\hss}}

\newcommand{\btc}{\begin{tabular}{c}}
\newcommand{\btbl}{\begin{tabular}{l}}
\newcommand{\et}{\end{tabular}}

\newcommand{\Nus}[1]{\mathrm{Nu}\!\left( #1 \right) }

\newcommand{\ssN}{{\scriptscriptstyle N}}

\newcommand{\rmno}{{\rm \bar{o}}} 
\newcommand{\rmo}{{\rm o}}

%% file: Koopman.bbl
\begin{thebibliography}{1}

\bibitem{hespanha2018linear}
Joao~P Hespanha.
\newblock {\em Linear systems theory}.
\newblock Princeton university press, 2018.

\bibitem{korda2018convergence}
Milan Korda and Igor Mezi{\'c}.
\newblock On convergence of extended dynamic mode decomposition to the koopman
  operator.
\newblock {\em Journal of Nonlinear Science}, 28(2):687--710, 2018.

\bibitem{mauroy2018global}
Alexandre Mauroy and Igor Mezi{\'c}.
\newblock Global computation of phase-amplitude reduction for limit-cycle
  dynamics.
\newblock {\em Chaos: An Interdisciplinary Journal of Nonlinear Science},
  28(7):073108, 2018.

\bibitem{schmid2010dynamic}
Peter~J Schmid.
\newblock Dynamic mode decomposition of numerical and experimental data.
\newblock {\em Journal of fluid mechanics}, 656:5--28, 2010.

\bibitem{williams2015data}
Matthew~O Williams, Ioannis~G Kevrekidis, and Clarence~W Rowley.
\newblock A data--driven approximation of the koopman operator: Extending
  dynamic mode decomposition.
\newblock {\em Journal of Nonlinear Science}, 25(6):1307--1346, 2015.

\end{thebibliography}
